\documentclass[epj
]{svjour}
\usepackage{graphics}
\usepackage{epsfig}
\usepackage{subfigure}

\usepackage{mathptmx}
\usepackage{amsfonts}
\usepackage{amsmath}
\usepackage{amssymb,bm}
\usepackage{bbm}
\usepackage{bigstrut}
\usepackage{fix-cm}
\usepackage{array}
\usepackage{enumerate}

\newcommand\bbone{\ensuremath{\mathbbm{1}}}

\renewcommand{\jmath}{j}

\begin{document}
\title{Ab initio-driven nuclear energy density functional method}
\subtitle{A proposal for safe/correlated/improvable parametrizations of the off-diagonal EDF kernels}
\author{T. Duguet\inst{1,2,3} \and M. Bender\inst{4,5}  \and J.-P. Ebran\inst{6} \and T. Lesinski\inst{1} \and V. Som\`a\inst{1}
}                     
\institute{
CEA, Centre de Saclay, IRFU/Service de Physique Nucl{\'e}aire, F-91191 Gif-sur-Yvette, France 
\and 
KU Leuven, Instituut voor Kern- en Stralingsfysica, 3001 Leuven, Belgium
\and 
National Superconducting Cyclotron Laboratory
             and Department of Physics and Astronomy,
             Michigan State University, East Lansing, MI 48824, USA
\and 
Universit\'e Bordeaux, Centre d'Etudes Nucl\'eaires de
            Bordeaux Gradignan, UMR5797, F-33175 Gradignan, France
\and 
CNRS/IN2P3, Centre d'Etudes Nucl\'eaires de Bordeaux
            Gradignan, UMR5797, F-33175 Gradignan, France
\and 
CEA, DAM, DIF, F-91297 Arpajon, France
}
%
%
\abstract{This programmatic paper lays down the possibility to reconcile the necessity to resum many-body correlations into the energy kernel with the fact that safe multi-reference energy density functional (EDF) calculations cannot be achieved whenever the Pauli principle is not strictly enforced, as is for example the case when many-body correlations are parametrized under the form of empirical density dependencies. Our proposal is to exploit a newly developed ab initio many-body formalism to guide the construction of safe, explicitly correlated and systematically improvable parametrizations of the {\it off-diagonal} energy and norm kernels that lie at the heart of the nuclear EDF method. The many-body formalism of interest relies on the concepts of symmetry breaking {\it and} restoration that have made the fortune of the nuclear EDF method and is, as such, amenable to this guidance. After elaborating on our proposal, we briefly outline the project we plan to execute in the years to come. 
} 
\maketitle
\section{Introduction}
\label{intro}

Nuclear ab initio methods combine state-of-the-art models of elementary interactions with controlled many-body expansion techniques. Confronting their results with experimental data, ab initio calculations provide a test of our understanding of the strong interaction between nucleons, at least as long as the error associated with the solving of the many-body Schr\"odinger equation is smaller than the residual distance to the data. In that sense, such methods are not meant to account for data at all cost but rather use the discrepancy with them to gauge the quality of the input. When both nuclear interaction models and the solving of the many-body Schr\"odinger equation reach a good enough accuracy, ab initio methods can provide controlled extrapolations to experimentally unknown regions. 

Traditionally limited to the lightest nuclei~\cite{nogga97,wiringa00,Navratil:2009ut}, ab initio many-body methods have been extended tremendously over the last ten years to medium-mass closed-shell nuclei~\cite{Barbieri:2000pg,Kowalski:2003hp,Hagen:2010gd,Tsukiyama:2010rj,Binder:2012mk,Cipollone:2013zma} and to those displaying an open-shell character~\cite{StolarczykMonkhorst,soma11a,Hergert:2013uja,Soma:2013xha,Signoracci:2014dia}. Based on this extended reach, ab initio many-body calculations are providing an even more stringent test of nuclear interactions~\cite{Binder:2013xaa,Soma:2013xha,Hergert:2014iaa} built within the frame of, e.g., chiral effective-field theory~\cite{valderrama12a,Epelbaum:2014efa} than they have done so far. 
 
The nuclear energy density functional (EDF) method provides a powerful quantum mechanical tool that can be applied to all bound atomic nuclei~\cite{bender03b}, irrespective of their mass and isospin, thanks to a low computational cost. The EDF method is thus characterized by an extended reach and aims, from the outset, at accounting for empirical phenomena with the highest possible precision, at least in the vicinity of the region where experimental data are available to adjust the parameters entering the EDF kernels. It is, however, an effective method that does not provide (direct) information about interactions between elementary nucleons. Most importantly, the empirical formulation of currently available parametrizations of the EDF kernels, i.e. the fact that they are not rooted into sound many-body methods, makes EDF predictions away from known data unreliable if not plagued with spuriousities leading to critical pathologies~\cite{dobaczewski07}. Those pathologies are the consequences of a violation of the Pauli exclusion principle that eventually contaminates state-of-the-art multi-reference EDF calculations with nonphysical contributions to the energy~\cite{Lacroix:2008rj,Bender:2008rn,Duguet:2008rr}. The violation of the Pauli principle in the EDF kernels relates itself to the traditional way of parametrizing many-body correlations under the form of density dependencies or to a relaxation of specific interrelations between the coefficients of the various terms at play. Solutions to better formulate the multi-reference EDF method and in particular the restoration of symmetries constitute a top priority today~\cite{duguet06a,duguet10a,duguet14a}.

\begin{figure}[t!]
\begin{center}
\includegraphics[clip=,width=0.5\textwidth,angle=0]{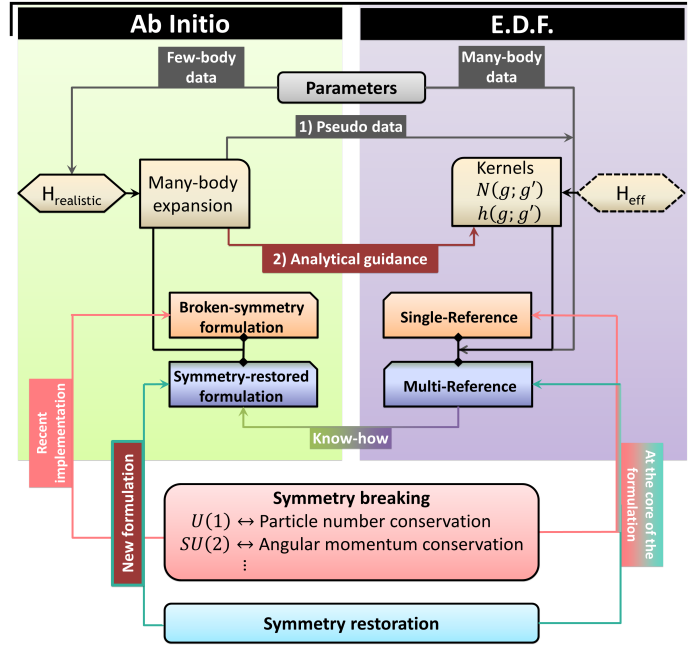} \\
\end{center}
\caption{
\label{connection}
(Color online) Potential connection and cross fertilization between EDF and ab initio many-body methods.}
\end{figure}

It is the goal of this programmatic document to propose one possible way to overcome these limitations by rooting the formulation of the nuclear EDF method into sound and appropriate many-body techniques. As sketched in Fig.~\ref{connection}, there are typically two ways how ab initio many-body methods can be used to improve on the current status of EDF calculations
\begin{enumerate}
\item When ab initio methods based on realistic inter-nucleon interactions are mature enough, their predictions for experimentally unknown nuclei can be used as pseudo-data to better constrain parameters entering the functional form of the nuclear EDF kernels. As of today, however, ab initio calculations of mid-mass nuclei, which constitute the natural overlap region with the EDF method, have not reached such a maturity yet.
\item The expansion of the Schr\"odinger equation at play in a given many-body method can be used as a mathematical guidance to build sound parametrizations of the EDF kernels that incorporate much needed correlations while avoiding unwanted pathologies. For this rationale to be operative, the many-body technique of interest {\it must} build on the same key concepts as those underlying the EDF method, i.e. {\it the breaking and the restoration of symmetries}. It is only in the last five years that ab initio many-body methods combining self-consistent Green's function (SCGF)~\cite{soma11a,Soma:2013xha} or coupled-cluster (CC)~\cite{StolarczykMonkhorst,henderson14a,Signoracci:2014dia} with the concept of symmetry breaking have been implemented to tackle open-shell nuclei. It is even more recently that methods adding the exact restoration of the broken symmetry have been formulated~\cite{duguet15a}.
\end{enumerate}
While the first point must be postponed to several years in the future, the present work elaborates on the second.
The proposal is fundamentally based on the novel many-body formalism, or more specifically on its simplified many-body perturbation theory (MBPT) version, proposed in Ref.~\cite{duguet15a}. As will be described below, this formalism consistently builds on the successive breaking and restoration of symmetries and is, as such, the first and only full-fledged many-body technique that can provide a constructive approach to {\it off-diagonal} EDF kernels.

The paper is organized as follows. Section~\ref{EDF} provides a brief account of the nuclear EDF formalism as it stands today. Basic ingredients, key concepts and current limitations are underlined to facilitate in Sec.~\ref{MBPT} the introduction of the novel many-body method used as a guidance to build the new family of EDF parametrizations. The extended EDF scheme that emerges from this proposal is outlined in Sec.~\ref{novelEDFscheme}. Conclusions and perspectives are given in Sec.~\ref{concluandperspect}.

\section{Modern nuclear EDF formalism}
\label{EDF}

We provide a brief introduction to the nuclear EDF formalism based on the more complete account given in Ref.~\cite{duguet14a}. As such, we do not aim at reviewing the status of the field, at covering all possible ramifications of the approach or at presenting recent achievements and applications. For standard reviews that cover the connection to empirical data, we refer the reader to, e.g., Refs.~\cite{bender03b,Niksic:2011sg}. 

\subsection{Reference states and off-diagonal EDF kernels}
\label{offdiagobasics}

The ingredients lying at the heart of the EDF method are the so-called {\it off-diagonal} norm $N(g';g)$ and energy $H(g';g)$ kernels~\cite{duguet14a}. The norm kernel is traditionally defined as the plain overlap
\begin{equation}
N(g';g) \equiv \langle \Phi(g')  | \Phi(g) \rangle \, ,  \label{offdiagnormkernel}
\end{equation}
between two many-body states of Bogoliubov type. The latter designates normalized\footnote{This corresponds to using the convention $N(g;g)=1$.} product states of the form
\begin{equation}
| \Phi(g) \rangle = {\cal C}\prod_{\mu} \beta^{(g)}_{\mu} \vert 0 \rangle \,\,\, ,
\label{Intro_met:product_state}
\end{equation}
where quasi-particle creation and annihilation operators satisfying $\{\beta^{(g)}_{\mu},\beta^{(g)\dagger}_{\nu} \} =\delta_{\mu\nu}$ relate to particle operators $\{c^{\dagger}_{\alpha} ; c_{\alpha}\}$ associated with a basis of the one-body Hilbert space ${\cal H}_1$ through the so-called Bogoliubov transformation
\begin{subequations}
\label{transfodetails}
\begin{align}
\beta^{(g)}_{\mu} &\equiv \sum_\alpha \Big( U^{(g)\ast}_{\alpha \mu} c_\alpha + V^{(g)\ast}_{\alpha \mu} c^\dagger_\alpha \Big) \,, \\
\beta^{(g)\dagger}_{\mu} &\equiv \sum_\alpha \Big( V^{(g)}_{\alpha \mu } c_\alpha  +  U^{(g)}_{\alpha \mu } c^\dagger_\alpha\Big) \,.
\end{align}
\end{subequations}
Matrices $U^{(g)}$  and $V^{(g)}$ make up the unitary Bogoliubov transformation~\cite{ring80a}. 

The collective index $g \equiv |g| e^{i  \Omega}$ labeling the many-body states gathers a set of order parameters characterizing the potential breaking of symmetries of the underlying nuclear Hamiltonian\footnote{In the most general setting, the label $g$ may also incorporate non-collective quantum numbers characterizing a set of quasi-particle excitations~\cite{ballythesis,Bally:2014jsa}.}. The norm $|g|$ of the order parameter tracks the extent to which $| \Phi(g) \rangle$ breaks the symmetry, i.e. its "deformation", whereas the phase $\Omega \equiv {\rm Arg}(g)$ characterizes the orientation of the deformed body with respect to the chosen reference frame\footnote{For certain symmetries, the compact notation $g \equiv |g| e^{i  \Omega}$ is schematic as the "phase" $\Omega$ actually collects several angles. See Sec.~\ref{limitationsu2} below for the case of $SU(2)$.}. For nuclei, the main symmetries to be considered are (i) $SU(2)$ associated with rotational invariance in real space and the conservation of angular momentum and (ii) $U(1)$ associated with rotational invariance in gauge space and the conservation of particle number. The associated compact Lie group is generically defined as ${\cal G}\equiv \{ R(\Omega); \Omega \in D_{{\cal G}}\}$ and the connection between states differing by their angles writes as 
\begin{equation}
| \Phi(|g|,\Omega)\rangle = R(\Omega) | \Phi(|g|,0)\rangle \, .
\end{equation}

In general, the off-diagonal energy kernel 
\begin{equation}
H(g';g) \equiv h(g';g) \, N(g';g) \, ,
\end{equation}
invokes a functional 
\begin{equation}
h(g';g) \equiv h[\mathbf{\rho}^{g'g}, \mathbf{\kappa}^{g'g}, \mathbf{\kappa}^{gg'  \ast}] \, 
\end{equation}
of the off-diagonal normal and anomalous one-body density matrices constructed from the two product states involved through
\begin{subequations}
\begin{eqnarray}
\rho_{ij}^{g'g} \equiv \frac{\langle \Phi(g') | c^{\dagger}_{j} c_{i}| \Phi(g) \rangle}
                {\langle \Phi(g') | \Phi(g)  \rangle}  \, , \\
\kappa_{ij}^{g'g} \equiv \frac{\langle \Phi(g') | c_{j} c_{i}| \Phi(g) \rangle}
                {\langle \Phi(g')  | \Phi(g)  \rangle}    \, , \\
\kappa^{g g' \ast}_{ij} \equiv \frac{\langle \Phi(g') | c^{\dagger}_{i} c^{\dagger}_{j}| \Phi(g) \rangle}
                {\langle \Phi(g')  | \Phi(g)  \rangle} \, .
\end{eqnarray}
\end{subequations}
The two reference states implicated $| \Phi(g) \rangle$ and $| \Phi(g') \rangle$ are a priori different and provide the kernels with their {\it off-diagonal} character. 

\subsection{SU(2) symmetry as an example}
\label{limitationsu2}

For presentation purposes\footnote{Everything discussed throughout the paper can be generalized to any (combination of) symmetry(ies) of interest without running into any fundamental difficulty.}, we limit the discussion throughout the rest of the paper to the $SU(2)\equiv \{R(\Omega), \Omega \in  D_{SU(2)}\}$ group, where $\Omega\equiv(\alpha,\beta,\gamma)$ actually embodies the three Euler angles whose domains of definition are 
\begin{equation}
D_{SU(2)} \equiv D_{\alpha} \times D_{\beta} \times D_{\gamma}  = [0,4\pi] \times [0,\pi] \times [0,2\pi] \, .
\end{equation}
We utilize the unitary representation of $SU(2)$ on Fock space given by  
\begin{equation}
R(\Omega)  =e^{-\frac{i}{\hbar}\alpha J_{z}}e^{-\frac{i}{\hbar}\beta J_{y}}e^{-\frac{i}{\hbar}\gamma J_{z}} \, ,
\end{equation}
where the three components of the angular momentum vector $\vec{J} = \sum_{n=1}^{\text{A}} \vec{j}(n)$ take the second-quantized form
\begin{equation}
J_{i} = \sum_{\alpha\beta} (j_{i})_{\alpha\beta}  \, c^{\dagger}_{\alpha} c_{\beta} \, ,
\end{equation}
with $i=x,y,z$ and $(j_{i})_{\alpha\beta} \equiv \langle 1: \alpha | j_{i} | 1: \beta \rangle$. Those one-body operators make up the Lie algebra
\begin{equation}
[J_{i},J_{j}]=\sum_{k} \epsilon_{ijk} i\hbar \, J_{k} \, , \label{Lieidentity}
\end{equation}
where $\epsilon_{ijk}$ denotes the Levi-Civita tensor. The Casimir operator of the group built from the infinitesimal generators through a non-degenerate invariant bilinear form is the total angular momentum
\begin{equation}
J^{2}  \equiv \sum_{i=x,y,z} J^2_{i} \, ,
\end{equation}
which is the sum of a one-body and a two-body term, respectively defined as
\begin{subequations}
\label{J21and2}
\begin{eqnarray}
J^{2}_{(1)} &\equiv& \sum_{n=1}^{\text{A}} j^2(n)  = \sum_{\alpha\beta} j^2_{\alpha\beta}  \, c^{\dagger}_{\alpha} c_{\beta} \, , \\
J^{2}_{(2)} &\equiv& \sum_{n\neq n'=1}^{\text{A}}  \vec{j}(n)\cdot \vec{j}(n')  = \frac{1}{2} \sum _{\alpha\beta\gamma\delta} j\!j_{\alpha\beta\gamma\delta} \, c^{\dagger}_{\alpha} c^{\dagger}_{\beta} c_{\delta} c_{\gamma}  \, .
\end{eqnarray}
\end{subequations}
Their (direct-product) matrix elements are given by
\begin{subequations}
\begin{eqnarray}
j^2_{\alpha\beta}  & \equiv& \langle 1: \alpha | j^2 | 1: \beta \rangle \nonumber \\
&=& \sum_{i=x,y,z} \langle 1: \alpha | j^2_{i} | 1: \beta \rangle \, , \\
j\!j_{\alpha\beta\gamma\delta} &\equiv& \langle 1: \alpha; 2: \beta | j\!j | 1: \gamma ; 2: \delta \rangle \nonumber \\
&=&  2 \sum_{i=x,y,z} \langle 1 :  \alpha | j_{i} | 1: \gamma \rangle  \, \langle 2: \beta | j_{i} | 2: \delta \rangle  \, ,
\end{eqnarray}
\end{subequations}
from which antisymmetrized matrix elements are obtained through $\overline{\jmath \jmath}_{\alpha\beta\gamma\delta} \equiv j\!j_{\alpha\beta\gamma\delta} - j\!j_{\alpha\beta\delta\gamma}$.

Matrix elements of the irreducible representations (IRREPs) of $SU(2)$ are given by the so-called Wigner $D$-functions~\cite{varshalovich88a}
\begin{equation}
\langle \Psi^{JM}_{\mu} |  R(\Omega)  |\Psi^{J'K}_{\mu'} \rangle \equiv \delta_{\mu\mu'} \delta_{JJ'} D_{MK}^{J}(\Omega) \, ,
\end{equation}
where $| \Psi^{JM}_{\mu} \rangle$ is an eigenstate of $J^2$ and $J_{z}$
\begin{subequations}
\label{eigenequationJ}
\begin{eqnarray}
J^2 | \Psi^{JM}_{\mu} \rangle &=&  J(J+1)\hbar^2 | \Psi^{JM}_{\mu} \rangle \,\,\, , \\
J_{z} | \Psi^{JM}_{\mu} \rangle &=&  M\hbar | \Psi^{JM}_{\mu} \rangle \,\,\, .
\end{eqnarray}
\end{subequations}
with $2J \in \mathbb{N}$, $2M\in \mathbb{Z}$, $J-M\in \mathbb{N}$ and $-J\leq M \leq +J$. The $(2J\!+\!1)$-dimensional IRREPs are labeled by $J$ and are spanned by the $\{| \Psi^{JM}_{\mu} \rangle\}$ for fixed $J$ and $\mu$.

The volume of the group is
\begin{eqnarray}
v_{SU(2)} &\equiv& \int_{D_{SU(2)}} \!\!\! d\Omega = 16\pi^2 \, ,
\end{eqnarray}
such that the orthogonality of Wigner $D$-functions reads
\begin{equation}
\int_{D_{SU(2)}} \hspace{-0.7cm} d\Omega \, D_{MK}^{J \, \ast}(\Omega) \, D_{M'K'}^{J'}(\Omega)  =\frac{16\pi^{2}}{2J+1}\delta_{JJ'}\delta_{MM'}\delta_{KK'} \, . \label{orthogonality}
\end{equation}%

Focusing on the restoration of $SU(2)$ symmetry, it is unnecessary to consider the breaking of $U(1)$ symmetry in the first place. We thus limit ourselves for simplicity to (normalized) reference states of the Slater determinant type , i.e. to the case where the Bogoliubov reference states defined through Eqs. ~\ref{Intro_met:product_state}-\ref{transfodetails} reduce to
\begin{equation}
|  \Phi(g) \rangle \equiv \prod_{i=1}^{A} a_{i}^{(g) \dagger} \, | 0 \rangle \, ,
\end{equation}
such that the energy kernel depends solely on the normal density matrix in this case, i.e. $h(g';g) \equiv h[\mathbf{\rho}^{g'g}]$. The $A$ occupied (hole) orbitals in $|  \Phi(g) \rangle$ are labeled by $(i,j,k,l\ldots)$ while unoccupied (particle) orbitals are labeled by $(a,b,c,d\ldots)$. Greek labels $(\alpha,\beta,\gamma,\delta\ldots)$ represent occupied or unoccupied states indifferently. 

\subsection{SR and MR implementations of the EDF method}
\label{implementations}

The EDF method is embodied in two successive levels of implementation. The first step makes use of the sole {\it diagonal} part of the kernels. The energy kernel  $H(g;g) = h(g;g)$ at play involves one symmetry-breaking state at a time such that this implementation is denoted as the single-reference (SR) level. The reference state and the energy are obtained by minimizing the latter under the constraint that the magnitude of the order parameter is fixed to a given value $|g|$, i.e. the SR energy is given by $E_{{\rm SR}}^{|g|} \equiv {\rm Min}_{\{| \Phi(g) \rangle\}}  \Big\{ {\cal E}_{|g|} \Big\}$ along with
\begin{equation}
{\cal E}_{|g|} \equiv h(g;g) - \lambda_{|g|} \, \Big[|g| - |\langle \Phi(g) | G | \Phi(g) \rangle |\Big]  \, ,
\end{equation}
where $G$ denotes the operator characterizing the order parameter, i.e. an operator whose average value is zero in a symmetry-conserving state. The minimization of ${\cal E}_{|g|} $ leads to solving equations of motion of the form
\begin{equation}
{\rm h}^{(g)} \, \varphi^{(g)}_{\alpha}  = \epsilon^{|g|}_{\alpha}  \, \varphi^{(g)}_{\alpha} \, ,
\end{equation}
where the diagonal one-body field
\begin{equation}
{\rm h}^{(g)}   \equiv  \frac{\delta {\cal E}_{|g|}}{\delta \rho^{gg \ast}} 
\end{equation}
is obtained as the functional derivative of the diagonal kernel with respect to the diagonal density matrix. The latter is characterized by $\rho^{gg}_{\alpha\beta} = n^{|g|}_\alpha \, \delta_{\alpha\beta}$, where $n^{|g|}_i = 1$ for hole states and $n^{|g|}_a = 0$ for particle states. The independence of the energy with respect to $\Omega$ denotes the existence of a pseudo-Goldstone mode (whenever $|g|\neq 0$) and relates to the independence of the kernels under a simultaneous rotation of both left and right states about the same angle~\cite{Robledo07a}. This property indicates that off-diagonal kernels must only depend on the {\it difference} of the angles defining the left and right states, i.e. 
\begin{subequations}
\begin{eqnarray}
N(|g'|,\Omega';|g|,\Omega) &=&  N(|g'|,0;|g|,\Omega-\Omega') \nonumber \\
&\equiv& N^{|g'||g|}(\Omega-\Omega') \, , \\
h(|g'|,\Omega';|g|,\Omega) &=&  h(|g'|,0;|g|,\Omega-\Omega') \nonumber \\
&\equiv& h^{|g'||g|}(\Omega-\Omega') \, .
\end{eqnarray}
\end{subequations}
The SR description provides a first account of ground-state properties along with a selected set of spectroscopic information~\cite{bender03b}. Allowing the reference state to break symmetries of the underlying Hamiltonian is key to incorporate static collective correlations and thus to address all nuclei, irrespective of their closed- or open-shell character. 

The second step makes a full use of the {\it off-diagonal} kernels constructed from all pairs of reference states belonging to a given set and is thus denoted as the multi-reference (MR) level. In state-of-the-art calculations, the set may contain up to about $10^9$ different symmetry-breaking product states. A key feature of the MR description relates to the inclusion of quantum collective fluctuations. In particular, mixing states spanning $D_{{\cal G}}$ amounts to restoring the symmetry that was possibly broken at the SR level. Focusing on such a feature\footnote{The magnitude of the order parameter $|g|$, i.e. the "deformation", is thus omitted throughout the rest of the paper. We refer the reader to Ref.~\cite{duguet14a} for an account of the MR formalism that includes the treatment of collective fluctuations associated with $|g|$.} and sticking to $SU(2)$, the MR energy associated with a state carrying good angular momentum reads as
\begin{eqnarray}
E^{J}_{{\rm MR}} &\equiv& \frac{\sum_{K'K} f^{J\ast}_{K'} \, f^{J}_K  \int_{D_{SU(2)}} \! d\Omega \, D^{J \, \ast}_{K'K}(\Omega) \,h(\Omega)\, N(\Omega)}{\sum_{K'K} f^{J\ast}_{K'} \, f^{J}_K  \int_{D_{SU(2)}} \! d\Omega \, D^{J \, \ast}_{K'K}(\Omega) \, N(\Omega)} \, . \label{EJ}
\end{eqnarray}
To better appreciate the content of Eq.~\ref{EJ}, one must realize that, as functions defined on $D_{SU(2)}$, off-diagonal norm and energy kernels can be expanded over the IRREPs of the group according to\footnote{When treating the restoration of neutron or proton numbers associated with the $U(1)$ group, the corresponding expansion is nothing but the Fourier expansion.}
\begin{subequations}
\label{expansion}
\begin{eqnarray}
N(\Omega) & \equiv& \sum_{JK'K} \, \textmd{N}^{J}_{K'K} \, D^{J}_{K'K}(\Omega) \, , \\
h(\Omega) \, N(\Omega)  & \equiv& \sum_{JK'K} \, \textmd{E}^{J}_{K'K}\, \textmd{N}^{J}_{K'K} \, D^{J}_{K'K}(\Omega) \, .
\end{eqnarray}
\end{subequations}
Exploiting the orthogonality of the IRREPs (Eq.~\ref{orthogonality}), the symmetry-restored energy can be written as
\begin{eqnarray}
E^{J}_{{\rm MR}} &=& \frac{\sum_{K'K} f^{J\ast}_{K'} \, f^{J}_K  \, \textmd{E}^{J}_{K'K} \, \textmd{N}^{J}_{K'K} }{\sum_{K'K} f^{J\ast}_{K'} \, f^{J}_K \, \textmd{N}^{J}_{K'K}  }  \, , \label{rewriteEJ}
\end{eqnarray}
and is nothing but a normalized mixing of the coefficients appearing in expansion~\ref{expansion}. In Eqs.~\ref{EJ} and~\ref{rewriteEJ}, the sum over $(K',K)$ mixes the components of the targeted IRREP to remove the nonphysical dependence on the orientation of the deformed reference state. The coefficients of the mixing $f^{J}_K$ are generally unknown and are typically determined utilizing the fact that the ground-state energy is a variational minimum. This eventually leads to solving a Hill-Wheeler-Griffin equation~\cite{Hill53,ring80a,bender03b}
\begin{equation}
\sum_{K=-J}^{+J}  \left(\textmd{E}^{J}_{K'K} -  E^{J}_{{\rm MR}} \right)\textmd{N}^{J}_{K'K} \, f^{J}_K = 0 \, . \label{HWG}
\end{equation}
Eventually, the MR description not only refines properties already computed at the SR level but also enlarges the number of accessible characteristics of the system, especially by reliably addressing observables that intimately depend on the fulfillment of symmetry selection rules, e.g. electromagnetic transitions of specific multipolarity.  All in all, the concept of symmetry breaking and restoration is at the heart of the nuclear EDF method as it constitutes a powerful tool to include collective correlations that are otherwise extremely costly to grasp within a symmetry conserving approach.

\subsection{Explicit formulation of the kernels}
\label{implementations}

Working under the hypothesis that left and right states solely differ by a rotation $R(\Omega)$, we write 
\begin{subequations}
\begin{eqnarray}
|  \Phi (0) \rangle &\equiv&  \prod_{i=1}^{A} a_{i}^{\dagger} \, |  0 \rangle \, , \\
|  \Phi (\Omega) \rangle &\equiv& \prod_{i=1}^{A} a_{\bar{\imath}}^{\dagger} \, |  0 \rangle \, .
\end{eqnarray}
\end{subequations}
where rotated orbitals are defined from unrotated ones through
\begin{subequations}
\begin{eqnarray}
| \bar{\alpha} \rangle &\equiv& R(\Omega)  | \alpha \rangle = \sum_{\beta} R_{\beta\alpha}(\Omega)  | \beta \rangle \, , \\
 a_{\bar{\alpha}}^{\dagger} &\equiv& R(\Omega) \, a^{\dagger}_{\alpha} \, R^{\dagger}(\Omega) = \sum_{\beta} R_{\beta\alpha}(\Omega)  a_{\beta}^{\dagger} \, ,
\end{eqnarray}
\end{subequations}
with $R_{\alpha\beta}(\Omega) \equiv \langle \alpha | R(\Omega) | \beta \rangle $ the unitary transformation matrix connecting the rotated basis to the unrotated one. The overlap between the two Slater determinants $|  \Phi(0) \rangle$ and $|  \Phi (\Omega) \rangle$ can be expressed as~\cite{blaizot86}
\begin{eqnarray}
\langle \Phi (0) | \Phi(\Omega) \rangle &=& \text{det} \, M(\Omega) \, , \label{kernel}
\end{eqnarray}
where $M_{ij}(\Omega)$ is the $A\times A$ reduction of $R_{\alpha\beta}(\Omega)$ to the subspace of hole states of $| \Phi (0) \rangle$. 

As Eq.~\ref{offdiagnormkernel} testifies, the off-diagonal norm kernel is explicitly given as the plain overlap between the two reference states involved. Formulated phenomenologically, the energy kernel is either taken as the off-diagonal matrix element of an effective Hamilton operator\footnote{In the present context, the wording {\it effective} forbids the possibility that the Hamiltonian depends on the many-body solution via, e.g., a dependence on the (off-diagonal) density (matrix) of the system.} or as a more general functional of the off-diagonal density matrices. In the former case, it takes the form of a low-order polynomial functional of the density matrices with specific inter-relations between its coefficients. In the latter case the functional typically involves further dependencies on non-integer powers of the off-diagonal density matrices and less constrained inter-relations between the coefficients involved. 

In practice, all modern parametrizations characterized by a good enough performance (irrespective of the merit function used to judge this performance) do not fall in the category of functionals formulated as the strict off-diagonal matrix element of an effective Hamilton operator. As a result, MR calculations, although characterized by an apparent success throughout the first years of their applications, have been shown to be plagued with critical pathologies~\cite{dobaczewski07}. As of today, the consensus among practitioners is that those pathologies must not be overlooked and require a decisive solution~\cite{ESNTworkshopnov2014}. Two possible options are in sight; i.e. (i) regularize the pathologies a posteriori for a given functional or (ii) limits one-self to energy kernels that are strictly based on an effective Hamiltonian operator. While the former route has been the first one followed~\cite{Lacroix:2008rj,Bender:2008rn,Duguet:2008rr} and still offers some opportunities~\cite{Satula:2014nba}, the latter is now becoming popular~\cite{sadoudi11thesis,Sadoudi:2012jg,Dobaczewski:2012cv,sadoudi13a,Bennaceur:2013fua,lacroix14a}. 

By construction, off-diagonal energy kernels strictly based on an effective Hamilton operator are free from any pathology and thus {\it safe} to be used meaningfully in MR calculations. Such a scheme starts by introducing a tractable effective Hamiltonian operator typically containing, e.g., (simple) two and three-body pseudo potentials
\begin{subequations}
\label{pseudoH}
\begin{eqnarray}
H_{\text{eff}} &\equiv& T+V_{\text{eff}}+W_{\text{eff}}\\
&=& \sum_{\alpha\beta} t_{\alpha\beta} \, c^\dagger_\alpha c_\beta^{\,} \\
&+&
\left(\frac{1}{2!}\right)^{\! 2} \sum_{\alpha\beta\gamma\delta} \bar{v}^{\,\text{eff}}_{\alpha\beta\gamma\delta} \, c^\dagger_\alpha c^\dagger_\beta c_\delta^{\,} c_\gamma^{\,} \\
&+&
\left(\frac{1}{3!}\right)^{\! 2} \sum_{\alpha\beta\gamma\delta\epsilon\zeta} \bar{w}^{\,\text{eff}}_{\alpha\beta\gamma\delta\epsilon\zeta} \, c^\dagger_\alpha c^\dagger_\beta c^\dagger_\gamma c_\zeta^{\,} c_\epsilon^{\,} c_\delta^{\,} \, ,
\end{eqnarray}
\end{subequations}
where $t_{\alpha\beta}$ denotes matrix elements of the one-body kinetic energy operator whereas $\bar{v}^{\,\text{eff}}_{\alpha\beta\gamma\delta}$ and $\bar{w}^{\,\text{eff}}_{\alpha\beta\gamma\delta\epsilon\zeta}$ represent antisymmetrized matrix elements of two- and three-body pseudo potential operators, respectively. Taking the straight matrix element (i.e. mean-field approximation) of $H_{\text{eff}}$, the off-diagonal energy kernel reads, by virtue of the generalized, i.e. off-diagonal, Wick's theorem (GWT)~\cite{balian69a}, as
\begin{subequations}
\label{Hbasedkernel}
\begin{eqnarray}
h(\Omega)  &\equiv& \frac{\langle \Phi(0) | H_{\text{eff}} | \Phi(\Omega) \rangle}{\langle \Phi(0) | \Phi(\Omega) \rangle} \label{Hbasedkernel1} \\
  &=&  \,\, \sum_{\alpha\beta} t_{\alpha\beta} \, \rho^{0\Omega}_{\beta\alpha} \label{Hbasedkernel2} \\
&+& 
\frac{1}{2} \sum_{\alpha\beta\gamma\delta} \bar{v}^{\,\text{eff}}_{\alpha\beta\gamma\delta}  \, \rho^{0\Omega}_{\gamma\alpha} \, \rho^{0\Omega}_{\delta\beta}
\label{Hbasedkernel3} \\
&+&
\frac{1}{6} \sum_{\alpha\beta\gamma\delta\epsilon\zeta} \bar{w}^{\,\text{eff}}_{\alpha\beta\gamma\delta\epsilon\zeta}   \, \rho^{0\Omega}_{\delta\alpha} \, \rho^{0\Omega}_{\epsilon\beta} \, \rho^{0\Omega}_{\zeta\gamma} \, .
\end{eqnarray}
\end{subequations}
Consequently, $h(\Omega)$ takes the form of a functional of the off-diagonal normal density matrix\footnote{When considering general reference states of the Bogoliubov type, additional terms depending on the anomalous off-diagonal density matrix are generated from the same pseudo-potentials~\cite{duguet14a}.} containing linear, bilinear and trilinear terms\footnote{This can obviously be extended to a quadrilinear terms by using a four-body pseudo-potential etc.} with specific interrelations between the coefficients of the polynomial.

Energy kernels of the form given by Eqs. \ref{pseudoH}-\ref{Hbasedkernel} and based on extended Skyrme~\cite{sadoudi13a} and/or finite-range~\cite{Dobaczewski:2012cv,lacroix14a} pseudo-potentials (without any density-dependent coupling) are currently being constructed and implemented. Although it is too early to declare success or failure, it already appears that the restrictive definition of the energy kernel may result in a lack of flexibility~\cite{Sadoudi:2012jg,ESNTworkshopnov2014}. As a matter of fact, and as can be expected from the pioneering work carried out in the 1970's~\cite{sadoudi13a}, it is a challenge to describe quantitatively all the desired phenomenology\footnote{This typically includes empirical characteristics of the nuclear equation of state, nuclear ground-states observables, including pairing properties, along with gross spectroscopic features of a (large) set of nuclei.} on the basis of an uncorrelated energy kernel (Eq. \ref{Hbasedkernel1}), independently of the (tractable) form of $H_{\text{eff}}$\footnote{In fact, this difficulty is what triggered the use of density dependencies that is now the standard in almost all modern parametrizations of the energy kernel.}. Of course, this difficulty must be put in perspective with the invaluable capacity to perform safe MR calculations with such parametrizations of the energy kernel. If this limitation happens to be too significant, one must question the routes that remain to be followed. In doing so, one can typically think of
\begin{enumerate}
\item Adding higher-order many-body terms to $H_{\text{eff}}$ to simulate missing correlations in the kernel,
\item Abandoning MR calculations and come back to general empirical parametrizations of the diagonal energy kernel.
\end{enumerate}
It is not clear how to proceed systematically with the first solution that might anyway become quickly impractical. As for the second solution, it is not something that one should be willing to follow yet. Indeed, the potentiality of state-of-the-art MR codes is too fantastic\footnote{See e.g. Ref.~\cite{Bally:2014jsa} for the recent development of MR calculations of odd nuclei or Ref.~\cite{satulaisospinmixing} for the computation of isospin-symmetry-breaking effects in superallowed Fermi beta decay in view of testing the unitarity of the Cabibbo-Kobayashi-Maskawa flavour-mixing matrix.} to give up on them at this point in time. The goal of the present document is to propose an alternative route that combines two wanted features, i.e. an approach that
\begin{enumerate}
\item Follows the strict Hamiltonian-based method and thus leads to safe MR calculations by construction,
\item Provides a way to encode missing correlations into the off-diagonal kernel(s) in a controlled and systematic fashion. 
\end{enumerate}
As such, the strategy is to balance the complexity between the effective Hamiltonian used on the one hand and the functional form of the kernels on the other hand. This is done by building  off-diagonal norm and energy kernels {\it beyond} the uncorrelated form given, respectively, by Eqs.~\ref{offdiagnormkernel} and~\ref{Hbasedkernel1} through a consistent expansion of {\it fully correlated} off-diagonal many-body kernels. 

\section{MBPT of off-diagonal kernels}
\label{MBPT}

Building on an earlier work~\cite{duguet03a}, ab initio MBPT and CC theories based on a symmetry-breaking reference state have been generalized in Ref.~\cite{duguet15a} in such a way that the symmetry is {\it exactly and consistently restored at any truncation order}. While Ref.~\cite{duguet15a} focused on the (breaking and the) restoration of $SU(2)$ symmetry, the case of U(1) symmetry has been formulated even more recently~\cite{duguet15b}. This set of novel many-body theories rely on the expansion of fully correlated {\it off-diagonal norm and energy kernels} and are, as such, perfectly suited to guide the systematic construction of their {\it effective} EDF counterparts. 

Results obtained in Ref.~\cite{duguet15a} are recalled here in the simple case of second-order MBPT. 
In view of their use as a basis for a novel EDF scheme, we make explicit the effective character of the Hamiltonian, $H_{\text{eff}}$, in terms of which the new EDF kernels are expressed. The present discussion is limited to a two-body pseudo potential $V_{\text{eff}}$, i.e. multi-body operators of higher rank in $H_{\text{eff}}$ are discarded for now. Derivations are omitted and the interested reader is referred to Ref.~\cite{duguet15a} for technical details. 

\subsection{Fully correlated off-diagonal kernels}
\label{exactkernels}

We introduce the evolution operator\footnote{Reference~\cite{duguet15a} being dedicated to {\it ab initio} calculations, the formalism is formulated having a realistic nuclear Hamiltonian $H$ in mind. This does not however prevent one from applying the same mathematical formalism in terms of an {\it effective} Hamiltonian $H_{\text{eff}}$. See Sec.~\ref{Remarks} for further comments on the connection between  both implementations.} in {\it imaginary} time\footnote{The time $\tau$ is real and given in units of MeV$^{-1}$.} 
\begin{eqnarray}
{\cal U}(\tau) &\equiv& e^{-\tau H_{\text{eff}}}  \label{evoloperator}
\end{eqnarray}
and the time-evolved many-body state
\begin{eqnarray}
| \Psi (\tau) \rangle &\equiv& {\cal U}(\tau) | \Phi (0) \rangle  \, , \label{evolstate} 
\end{eqnarray}
which satisfies the time-dependent Schr\"odinger equation
\begin{equation}
H_{\text{eff}} \, | \Psi (\tau) \rangle = -\partial_{\tau} | \Psi (\tau) \rangle  \, . \label{schroedinger}
\end{equation}
Having $| \Psi (\tau) \rangle$ at hand, we introduce a set of fully correlated, off-diagonal and time-dependent kernels
\begin{subequations}
\label{defkernels}
\begin{eqnarray}
N(\tau,\Omega) &\equiv& \langle \Psi (\tau)  | \bbone | \Phi(\Omega) \rangle  \, , \label{defnormkernel} \\
H(\tau,\Omega) &\equiv& \langle \Psi (\tau) | H_{\text{eff}} | \Phi(\Omega) \rangle  \, , \label{defenergykernel} \\
J_i(\tau,\Omega) &\equiv& \langle \Psi (\tau)  | J_i | \Phi(\Omega) \rangle  \, , \label{defJzkernel} \\
J^2(\tau,\Omega) &\equiv& \langle \Psi (\tau) | J^2 | \Phi(\Omega) \rangle  \, , \label{defJ2kernel} 
\end{eqnarray}
\end{subequations}
which relate to norm, energy, angular momentum projections and total angular momentum kernels, respectively. The energy and total angular momentum kernels can be further split into their one- and two-body components according to
\begin{subequations}
\label{defkernels1and2body}
\begin{eqnarray}
H(\tau,\Omega) &=& T(\tau,\Omega)+V(\tau,\Omega) \, , \label{defHkernel} \\
J^2(\tau,\Omega) &=& J^2_{(1)}(\tau,\Omega)+J^2_{(2)}(\tau,\Omega) \, . \label{defJ2oneandtwokernel}
\end{eqnarray}
\end{subequations}
In the following, a generic operator is denoted as $O$ while its off-diagonal kernel is referred to as 
\begin{eqnarray}
O(\tau,\Omega) &\equiv& \langle \Psi (\tau)  | O | \Phi(\Omega) \rangle  \, ,
\end{eqnarray}
and is noted in the infinite time limit as $O(\infty,\Omega) \equiv O(\Omega)$. Additionally, use is made of the {\it reduced} kernel defined through
\begin{equation}
{\cal O}(\tau,\Omega) \equiv \frac{O(\tau,\Omega)}{N(\tau,0)} \, . \label{reducedkernels}
\end{equation}
The advantage of introducing reduced kernels relates to the possibility to work with {\it intermediate normalization} at rotation angle $\Omega=0$, i.e. with ${\cal N}(\tau,0) \equiv 1$ for all $\tau$.

One can show~\cite{duguet15a} that the exact lowest energy $E^{J}_0$ of a given IRREP is obtained via
\begin{eqnarray}
E^{J}_0 &=& \frac{\sum_{K'K} f^{J\ast}_{K'} \, f^{J}_K  \int_{D_{SU(2)}} \! d\Omega \, D^{J \, \ast}_{K'K}(\Omega) \, {\cal H}(\Omega)}{\sum_{K'K} f^{J\ast}_{K'} \, f^{J}_K  \int_{D_{SU(2)}} \! d\Omega \, D^{J \, \ast}_{K'K}(\Omega) \, {\cal N}(\Omega)}  \, , \label{projected_energy}
\end{eqnarray}
where ${\cal H}(\Omega)$ and ${\cal N}(\Omega)$ denote the {\it fully correlated} (reduced) many-body kernels. From this point on, the challenge is to develop a many-body methodology to expand these kernels around the reference state $| \Phi (0) \rangle$. How to do this is explained at length in Ref.~\cite{duguet15a} where both the MBPT and CC expansions of off-diagonal kernels are fully worked out. We now report on the results obtained from the MBPT expansion at second order that are relevant to our purpose.

\subsection{Unperturbed system}
\label{unperturbedsystem}

The Hamiltonian is split into a one-body part $H_{0}$ and a residual two-body part $H_1$
\begin{equation}
\label{split1}
H_{\text{eff}} \equiv H_{0} + H_{1} \, ,
\end{equation} 
such that $H_{0}\equiv T+U$ and $H_{1}\equiv V^{\text{eff}}-U$, where 
\begin{equation}
\label{U}
U \equiv \sum_{\alpha\beta} u_{\alpha\beta} c^\dagger_\alpha c_\beta^{\,} \, ,
\end{equation} 
is a one-body operator that remains to be specified. Typically, we will advocate to take $H_{0}$ as the the sum of the zero and one-body part obtained by normal-ordering $H_{\text{eff}}$ with respect to the state $|  \Phi (0) \rangle $ that minimizes the expectation value of $H_{\text{eff}}$ under the possible breaking of $SU(2)$ symmetry, i.e. solving deformed Hartree-Fock equations for $H_{\text{eff}}$ one obtains 
\begin{equation}
\label{UHF}
u^{{\rm HF}}_{\alpha\beta} = \sum_{\gamma\delta}  \bar{v}^{\,\text{eff}}_{\alpha\gamma\beta\delta}  \, \rho^{00}_{\delta\gamma}  \, .
\end{equation} 
Correspondingly, $H_1$ is the (symmetry-breaking) normal-ordered two-body part of $V^{\text{eff}}$. In this context, one can let $|  \Phi (0) \rangle $ break $SU(2)$ symmetry spontaneously or force it to do so by adding an appropriate Lagrange constraint. This means that the product state $|  \Phi (0) \rangle$ is potentially not an eigenstate of $J^2$ and spans several IRREPs of $SU(2)$.

The operator $H_{0}$ can be written in diagonal form in terms of its (deformed) one-body eigenstates
\begin{equation}
H_{0}\equiv \sum_{\alpha}(e_{\alpha}\!-\!\mu) a_{\alpha}^{\dagger} a_{\alpha} \label{hzero} \, ,
\end{equation}
where the chemical potential $\mu$ is introduced for convenience. Thus, $|  \Phi (0) \rangle$ satisfies
\begin{subequations}
\begin{eqnarray}
H_{0}\, |  \Phi (0) \rangle &=& \varepsilon_{0} \, |  \Phi (0) \rangle \, , \\
\varepsilon_{0} &=& \sum_{i=1}^{A} (e_{i}-\mu)  \label{phi} \, .
\end{eqnarray}
\end{subequations}
The (deformed) Slater determinant $|  \Phi (0) \rangle$ necessarily possesses a {\it closed-shell} character, i.e. there exists a finite energy gap between the fully occupied shells below the Fermi energy (chemical potential) and the unoccupied levels above. The chemical potential is chosen to lie in the energy gap separating particle and hole orbitals, i.e.
\begin{equation}
e_{a}>\mu\;\; \text{and}\;\;\;e_{i}<\mu\label{eph} \, .
\end{equation}
Excited eigenstates of $H_0$ are obtained as particle-hole excitations of $| \Phi (0) \rangle$ 
\begin{eqnarray}
| \Phi^{ab\ldots}_{ij\ldots} (0) \rangle &\equiv& a^{\dagger}_{a} \, a_{i} \, a^{\dagger}_{b} \, a_{j} \ldots  |  \Phi (0) \rangle \, , 
\end{eqnarray}
with the eigenenergy
\begin{subequations}
\begin{eqnarray}
H_{0}\,| \Phi^{ab\ldots}_{ij\ldots} (0) \rangle &=& (\varepsilon_0 +\varepsilon^{ab\ldots}_{ij\ldots}) \, | \Phi^{ab\ldots}_{ij\ldots} (0) \rangle \, , \\
\varepsilon^{ab\ldots}_{ij\ldots} &=& e_a\!+\! e_b\! +\!\ldots\! -\! e_i\! -\! e_j\! -\! \ldots \label{phiexcit} \, .
\end{eqnarray}
\end{subequations}

Having $| \Phi (0) \rangle$ at hand, the off-diagonal one-body density matrix can be written as~\cite{blaizot86}
\begin{equation}
\rho^{0\Omega}  = \sum_{ij=1}^{A} |  \bar{\imath} \rangle \, M_{ij}^{-1}(\Omega) \, \langle j |  \label{rhohh} \, ,
\end{equation}
such that it acquires the form
\begin{eqnarray}
\rho^{0\Omega} &=& \left(
\begin{array} {cc}
\bbone^{hh} & 0   \\
0 & 0
\end{array}
\right) + \left(
\begin{array} {cc}
0 & 0  \\
R(\Omega)M^{-1}(\Omega) & 0
\end{array}
\right) \nonumber \\
&\equiv& \rho^{00} + \wp^{0\Omega} \, , \label{transdens}
\end{eqnarray}
where $\bbone^{hh}$ is the identity operator on the hole subspace of the one-body Hilbert space. The genuinely $\Omega$-dependent part $\wp^{0\Omega}$, which only connects particle kets to hole bras, vanishes for $\Omega=0$, i.e. $\wp^{00}=0$. The above partitioning of the off-diagonal density matrix can be summarized by writing its matrix elements under the form 
\begin{eqnarray}
\rho^{0\Omega}_{\alpha\beta} &\equiv& n_{\alpha} \, \delta_{\alpha\beta} +  (1-n_{\alpha}) \, n_{\beta} \, \wp^{0\Omega}_{\alpha\beta} \, . \label{truc} 
\end{eqnarray}
Making particle and hole indices explicit, one obtains equivalently
\begin{subequations}
\label{contractionsrho}
\begin{eqnarray}
\rho^{0\Omega}_{a b} &=& 0 \, , \label{contractionsrho1} \\
\rho^{0\Omega}_{i b} &=& 0 \, , \label{contractionsrho2} \\
\rho^{0\Omega}_{ij} &=& \delta_{ij} \, , \label{contractionsrho3} \\
\rho^{0\Omega}_{aj} &=& \wp^{0\Omega}_{aj} \, . \label{contractionsrho4}
\end{eqnarray}
\end{subequations}

\subsection{Second-order expansion of the energy kernel}
\label{expansionenergykernel}

Expanding the evolution operator in powers of $H_{1}$ under the form
\begin{eqnarray}
{\cal U}(\tau) &=& e^{-\tau H_{0}} \, \textmd{T}e^{-\int_{0}^{\tau}dt H_{1}\left(t\right)} \, ,
\label{evol1}%
\end{eqnarray}
where $\textmd{T}$ denotes the time-ordering operator and where
\begin{equation}
H_{1}\left( \tau\right)  \equiv e^{\tau H_{0}}H_{1}e^{-\tau H_{0}}
\end{equation}
defines the perturbation in the interaction representation, one can expand $O(\Omega)$ in perturbation and represent it diagrammatically~\cite{duguet15a}. This can be achieved by virtue of the GWT and by generalizing the definition of diagonal unperturbed one-body propagators to off-diagonal ones
\begin{equation}
G^{0}_{\alpha\beta}(\tau_1,\tau_2 ; \Omega) \equiv \frac{\langle \Phi |  \textmd{T}[a_{\alpha}(\tau_1) a_{\beta}^{\dagger}(\tau_2)] | \Phi(\Omega) \rangle}{\langle \Phi | \Phi(\Omega) \rangle}   \, . \label{onebodyproptrans}
\end{equation}
A key result is the demonstration that the kernel associated with any operator factorizes according to 
\begin{eqnarray}
O(\Omega) &\equiv& o(\Omega)\,N(\Omega) \, , \label{factorization}
\end{eqnarray}
where $o(\Omega)$ denotes the sum of all {\it connected} vacuum-to-vacuum diagrams {\it linked} to the operator $O$. This writes for the energy kernel as 
\begin{eqnarray}
H(\Omega) &\equiv& h(\Omega)\,N(\Omega) \, , \label{factorizationH}
\end{eqnarray}
which thus defines the connected/linked energy kernel $h(\Omega)$. 

\begin{figure}[t!]
\begin{center}
\includegraphics[clip=,width=0.21\textwidth,angle=0]{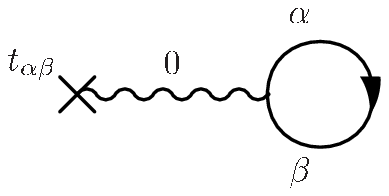}\\ \vspace{0.4cm}
\includegraphics[clip=,width=0.22\textwidth,angle=0]{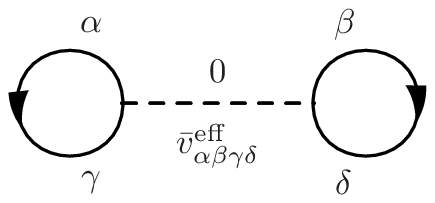}\\ \vspace{0.4cm}
\end{center}
\caption{
\label{diagramsTL}
First-order Feynman diagrams contributing to $h(\Omega)$. A full line carrying an arrow denotes a {\it off-diagonal} unperturbed one-body propagator $G^{0}_{\alpha\beta}(\tau_1,\tau_2 ; \Omega)$. See Ref.~\cite{duguet15a} for the details of the diagrammatic rules at play.}
\end{figure}

\begin{figure}[t!]
\begin{center}
\includegraphics[clip=,width=0.18\textwidth,angle=0]{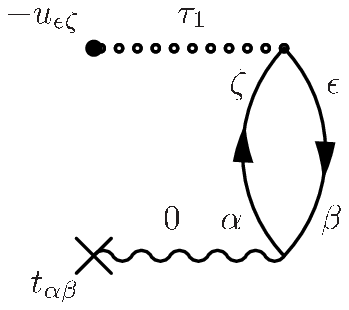}\\ \vspace{0.4cm}
\includegraphics[clip=,width=0.19\textwidth,angle=0]{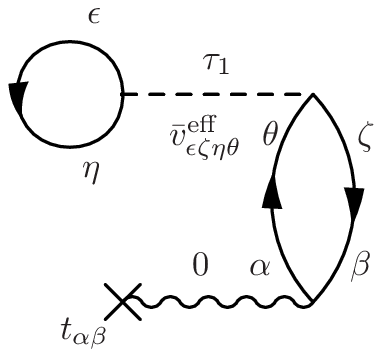}\\
\end{center}
\caption{
\label{diagramsTLbis}
Second-order Feynman diagrams contributing to $t(\Omega)$.}
\end{figure}

\begin{figure}[t!]
\begin{center}
\includegraphics[clip=,width=0.19\textwidth,angle=0]{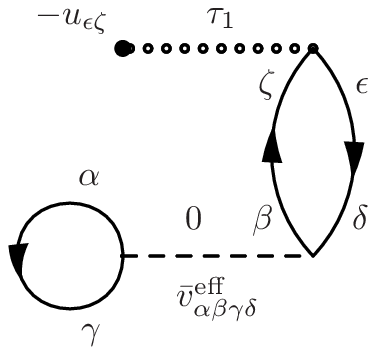}\\ \vspace{0.4cm}
\includegraphics[clip=,width=0.19\textwidth,angle=0]{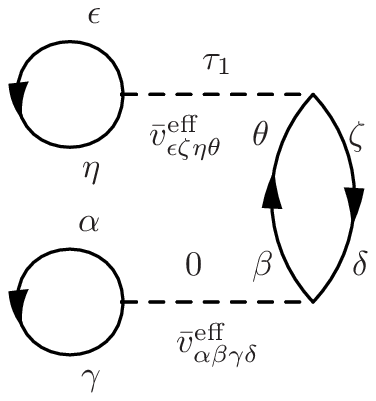}\\ \vspace{0.4cm}
\includegraphics[clip=,width=0.17\textwidth,angle=0]{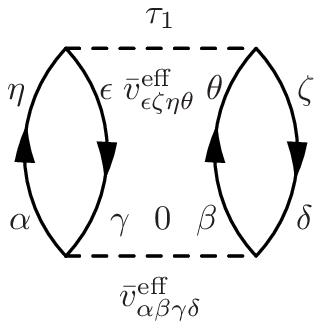}\\ 
\end{center}
\caption{
\label{diagramsVL}
Second-order Feynman diagrams contributing to $v(\Omega)$.}
\end{figure}

To first order\footnote{The counting of perturbative orders is shifted by one unit relative to the convention used in Ref~\cite{duguet15a}.} in MBPT, diagrams contributing to the connected/linked energy kernel are displayed in Fig.~\ref{diagramsTL} and read as
\begin{subequations}
\label{1storderh}
\begin{eqnarray} 
h^{(1)}(\Omega) &=& \frac{\langle \Phi (0) | H_{\text{eff}} | \Phi(\Omega) \rangle}{\langle \Phi (0) | \Phi(\Omega) \rangle}  \\
&=& \sum_{i} t_{ii}  + \sum_{ia} t_{ia} \, \wp^{0\Omega}_{ai}  \\
&+& \frac{1}{2}\sum_{ij} \bar{v}^{\,\text{eff}}_{ijij}  + \sum_{ijc} \bar{v}^{\,\text{eff}}_{ijcj} \, \wp^{0\Omega}_{ci} \nonumber  \\
&+& \frac{1}{2} \sum_{ijab} \bar{v}^{\,\text{eff}}_{ijab} \, \wp^{0\Omega}_{ai} \, \wp^{0\Omega}_{bj} \, ,
\end{eqnarray}
\end{subequations}
and is nothing but Eqs.~\ref{Hbasedkernel2}-\ref{Hbasedkernel3} in which particle and hole states have been specified and the off-diagonal density matrix has been split according to Eq.~\ref{transdens}. The expanded expression of the second-order contributions to $h(\Omega)$, which constitute the first corrections to the traditional effective mean-field kernel, is too cumbersome to be reported here and we refer the interested reader to Ref.~\cite{duguet15a}. This second-order correction to the {\it off-diagonal} energy kernel, whose diagrammatic representation is given in Figs.~\ref{diagramsTLbis} and~\ref{diagramsVL}, is indeed quite rich. To write it in a manageable form that is also more amenable to its numerical implementation, we first need to express $H_{\text{eff}}$ in convenient left and right bi-orthogonal single-particle bases that we now introduce.

The right basis $\{| \tilde{\alpha} \rangle\}$ is obtained by applying the non-unitary transformation
\begin{eqnarray}
B(\Omega) &\equiv& \bbone + \wp^{0\Omega} \, , \label{matrix}
\end{eqnarray}
onto the original basis $\{| \alpha \rangle\}$. Omitting for notational simplicity the explicit $\Omega$ dependence of the basis states thus obtained and separating original particle and hole states provides 
\begin{subequations}
\label{rightbasis}
\begin{eqnarray}
| \tilde{\imath} \rangle &=&  | i \rangle + \sum_{kc} | c \rangle \,  R_{ck}(\Omega) M^{-1}_{ki}(\Omega) \, , \label{rightbasis1} \\
| \tilde{a} \rangle &=&  | a \rangle \, . \label{rightbasis2}
\end{eqnarray}
\end{subequations}
Particle kets are thus left unchanged. The left basis  $\{\langle \tilde{\alpha} | \}$ is similarly obtained by applying the transformation
\begin{eqnarray}
B^{-1}(\Omega) &\equiv& \bbone - \wp^{0\Omega} \, , \label{inversematrix}
\end{eqnarray}
onto the original basis $\{\langle \alpha |\}$ such that
\begin{subequations}
\label{leftbasis}
\begin{eqnarray}
\langle \tilde{\jmath} | &=&  \langle j | \, , \label{leftbasis1} \\
\langle \tilde{b} | &=&  \langle b | - \sum_{kl} R_{bk}(\Omega) M^{-1}_{kl}(\Omega) \, \langle l |  \, . \label{leftbasis2}
\end{eqnarray}
\end{subequations}
Hole bras are thus left unchanged. Although we use for simplicity the same notation to characterize states in the left and right bases, the tilde is meant to underline their bi-orthogonal character. The latter, indicated by $\langle \tilde{\alpha} | \tilde{\beta} \rangle = \delta_{\alpha\beta}$ can be easily obtained from   $B^{-1}(\Omega)B(\Omega)=\bbone$. Given any n-body operator $O$, we introduce the transformed operator $\tilde{O}(\Omega)$ through
\begin{eqnarray}
\tilde{O}(\Omega) &\equiv& \left(\frac{1}{n!}\right)^2 \!\!\!\!\!\! \sum_{\alpha \ldots \beta\gamma\ldots\delta}\!\!\!\!\!\! O_{\tilde{\alpha}\ldots\tilde{\beta} \tilde{\gamma}\ldots\tilde{\delta}}(\Omega) \, a^{\dagger}_{\alpha} \ldots a^{\dagger}_{\beta} \, a_{\delta} \ldots a_{\gamma} \, , \label{deftransformedOp}
\end{eqnarray}
where creation and annihilation operators refer to the original eigenbasis $\{| \alpha \rangle\}$ of $H_0$ while left and right indices of the matrix elements refer to the associated bi-orthogonal system introduced above. With these definitions at hand, the first-order off-diagonal connected/linked energy kernel (Eq.~\ref{1storderh}) can be straightforwardly rewritten under the compact form
\begin{subequations}
\label{1storderhcompact}
\begin{eqnarray}
h^{(1)}(\Omega) &=& \sum_{i} t_{\tilde{\imath}\tilde{\imath}}(\Omega)  + \frac{1}{2}\sum_{ij} \bar{v}^{\,\text{eff}}_{\tilde{\imath}\tilde{\jmath}\tilde{\imath}\tilde{\jmath}}(\Omega) \label{1storderhcompact1} \\
&=& \langle \Phi (0) | \tilde{H}_{\text{eff}}(\Omega)  |\Phi (0) \rangle \, , \label{1storderhcompact2}
\end{eqnarray}
\end{subequations}
which represents the {\it diagonal} matrix element of the {\it transformed} Hamilton operator $\tilde{H}_{\text{eff}}(\Omega)$ at rotation angle $\Omega$.

The analytic expression of MBPT diagrams making up the connected/linked {\it off-diagonal} kernel of any operator $O$ can in fact be systematically reduced to the expression of the diagrams making its connected/linked {\it diagonal} kernel, at the sole price of involving the transformed operator $\tilde{O}(\Omega)$ expressed in the ($\Omega$-dependent) bi-orthogonal system~\cite{duguet15a}. Taking second-order MBPT as a prime example, and introducing for additional compactness the second-order expression of so-called one- and two-body cluster amplitudes\footnote{Choosing $U=U^{{\rm HF}}$ (Eq.~\ref{UHF}) cancels out the first contribution to ${\cal T}^{\dagger \, (2)}_{ia}(\Omega)$ (Eq.~\ref{perturbativeclusters1}).}
\begin{subequations}
\label{perturbativeclusters}
\begin{eqnarray}
{\cal T}^{\dagger \, (2)}_{ia}(\Omega) &\equiv&  - \frac{1}{e_a-e_i} \, \Big[\sum_{j}\bar{v}^{\,\text{eff}}_{i j a j} -  u_{ia}\Big] \label{perturbativeclusters1}  \\
&& - \sum_{jb} \frac{\bar{v}^{\,\text{eff}}_{i ja b}}{e_a+e_b-e_i-e_j} \, \wp^{0\Omega}_{bj} \, ,  \label{perturbativeclusters2} \\
{\cal T}^{\dagger \, (2)}_{ijab}(\Omega) &\equiv& - \frac{\bar{v}^{\,\text{eff}}_{ijab}}{e_a+e_b-e_i-e_j}  \, , \label{perturbativeclusters3}
\end{eqnarray}
\end{subequations}
one obtains the remarkable identity\footnote{The index $c$ indicates the {\it connected} character of the matrix element.}
\begin{subequations}
\label{remarkableidentity}
\begin{eqnarray}
o^{(2)}(\Omega) &=& \frac{\langle \Phi (0) | \Big[1+{\cal T}^{\dagger \, (2)}_{1}(\Omega)+{\cal T}^{\dagger \, (2)}_{2}(\Omega) \Big] O  |\Phi (\Omega) \rangle_{c}}{\langle \Phi (0) | \Phi (\Omega) \rangle} \nonumber \\
&=& \langle \Phi (0) | \Big[1+{\cal T}^{\dagger \, (2)}_{1}(\Omega)+{\cal T}^{\dagger \, (2)}_{2}(\Omega) \Big] \tilde{O}(\Omega)  |\Phi (0) \rangle_{c} \, , \nonumber
\end{eqnarray}
\end{subequations}
where the last matrix element can be worked out on the basis of the standard, i.e. diagonal, Wick theorem~\cite{wick50a}. Applying this to the energy kernel puts us in position to provide its complete expression at second order in MBPT under the compact form
\begin{subequations}
\label{full2ndh}
\begin{eqnarray}
h^{(2)}(\Omega) &= & \sum_{i} t_{\tilde{\imath}\tilde{\imath}}(\Omega)+\sum_{ia}  {\cal T}^{\dagger (2)}_{ia}(\Omega) \, t_{\tilde{a}\tilde{\imath}}(\Omega) \label{full2ndh1} \\
&+& \frac{1}{2}\sum_{ij} \bar{v}^{\,\text{eff}}_{\tilde{\imath}\tilde{\jmath}\tilde{\imath}\tilde{\jmath}}(\Omega)  + \sum_{ija}  {\cal T}^{\dagger (2)}_{ia}(\Omega) \, \bar{v}^{\,\text{eff}}_{\tilde{a}\tilde{\jmath}\tilde{\imath}\tilde{\jmath}}(\Omega)  \nonumber \\
&+& \frac{1}{4} \sum_{ijab}  {\cal T}^{\dagger (2)}_{ijab}(\Omega) \, \bar{v}^{\,\text{eff}}_{\tilde{a}\tilde{b}\tilde{\imath}\tilde{\jmath}}(\Omega) \, , \label{full2ndh2}
\end{eqnarray}
\end{subequations}
where Eqs.~\ref{full2ndh1} and~\ref{full2ndh2} denote the effective kinetic energy $t^{(2)}(\Omega)$ (Fig.~\ref{diagramsTLbis}) and potential energy $v^{(2)}(\Omega)$ (Fig.~\ref{diagramsVL}) parts, respectively. We wish to insist on the fact that, while Eq.~\ref{full2ndh} offers a compact and manageable expression of $h^{(2)}(\Omega)$, expanding the energy kernel fully would reveal its much richer content compared to its first order, i.e. effective mean-field, counterpart. 

\subsection{Consistent expansion of the norm kernel}
\label{expansionnormkernel}

So far, we have focused on the expansion of the connected/linked part of the kernel associated with an operator $O$. Dealing with the norm kernel amounts to taking $O=\bbone$, for which the connected/linked part is trivially equal to one such that one is left with the norm kernel itself (see Eq.~\ref{factorization}) by definition. A direct perturbative expansion of $N(\Omega)$ is possible but does not constitute the appropriate approach in the present context~\cite{duguet15a}. The consistent determination of the norm kernel relies on a specific argument, i.e. on the fact that applying the symmetry restoration scheme to the Casimir ($O=J^2$) of $SU(2)$ and to an infinitesimal generator ($O=J_z$) that commutes with it must give the expected values (respectively $J(J+1)\hbar^2$ and $M\hbar$) {\it independently of the order at which the expansion is truncated}. As demonstrated in Ref.~\cite{duguet15a}, this fundamental feature requires the reduced norm kernel ${\cal N}(\Omega)$ to satisfy three coupled ordinary differential equations  
\begin{subequations}
\label{NormkernelODE}
\begin{eqnarray}
\frac{\partial}{\partial \alpha} \, {\cal N}(\Omega) &=&  - \frac{i}{\hbar} \, j_z(\Omega) \, {\cal N}(\Omega) \, ,\label{NormkernelODE1} \\
\frac{\partial}{\partial \beta} \, {\cal N}(\Omega) &=& + \frac{i}{\hbar} \sin (\alpha) \, j_x(\Omega) \, {\cal N}(\Omega) \nonumber \\
&&- \frac{i}{\hbar}\cos (\alpha) \, j_y(\Omega) \, {\cal N}(\Omega)  \, ,\label{NormkernelODE2} \\
\frac{\partial}{\partial \gamma} \, {\cal N}(\Omega)  &=& - \frac{i}{\hbar} \sin (\beta) \, \cos (\alpha) \, j_x(\Omega) \, {\cal N}(\Omega)  \nonumber \\
&& - \frac{i}{\hbar} \sin (\beta) \, \sin (\alpha) \, j_y(\Omega) \, {\cal N}(\Omega)  \nonumber \\
&& - \frac{i}{\hbar} \cos (\beta) \, j_z(\Omega) \, {\cal N}(\Omega) \, ,\label{NormkernelODE3} 
\end{eqnarray}
\end{subequations}
where $\{j_x(\Omega),j_y(\Omega),j_z(\Omega)\}$ denote the connected/linked kernels of the three infinitesimal generators  $\{J_x, J_y, J_z)\}$ of $SU(2)$ and whose initial  condition is given by the intermediate normalization condition ${\cal N}(0)=1$. Working at a given MBPT order $n$, one must thus compute $\{j^{(n)}_x(\Omega),j^{(n)}_y(\Omega),j^{(n)}_z(\Omega)\}$ and solve Eq.~\ref{NormkernelODE} to determine ${\cal N}^{(n)}(\Omega)$. At second-order, this amounts to computing
\begin{eqnarray}
j^{(2)}_k(\Omega) &=& \sum_{i} (j_k)_{\tilde{\imath}\tilde{\imath}}(\Omega)+\sum_{ia}  {\cal T}^{\dagger (2)}_{ia}(\Omega) \, (j_k)_{\tilde{a}\tilde{\imath}}(\Omega)  \, \label{2ndorderjk}
\end{eqnarray}
and solving Eq.~\ref{NormkernelODE} to obtain consistently
\begin{eqnarray}
{\cal N}^{(2)}(\Omega) &\equiv& \aleph^{(2)}(\Omega) \, \langle \Phi (0) | \Phi(\Omega) \rangle \, . \label{2ndorderN}
\end{eqnarray}
In Eq.~\ref{2ndorderN}, $\aleph^{(n)}(\Omega)$ refers to the part of the norm kernel that factorizes in front of the plain overlap. If further limiting Eq.~\ref{2ndorderjk} to first order, Equation~\ref{NormkernelODE} reduces to the ODEs known~\cite{Har82a,Ena99a} to be fulfilled by the plain mean-field overlap, i.e. $\aleph^{(1)}(\Omega)=1$ and ${\cal N}^{(1)}(\Omega) = \langle \Phi (0) | \Phi(\Omega) \rangle$. Eq.~\ref{NormkernelODE} thus constitutes the proper generalization, to any order in the many-body expansion, of the ODEs known to be fulfilled by the uncorrelated norm kernel.

\subsection{Diagonal kernels}
\label{diagonal}

The many-body scheme developed in Ref.~\cite{duguet15a} for off-diagonal kernels provides a safe constructive ansatz and avoids the dangers of building off-diagonal kernels as {\it empirical} extension of known diagonal kernels. Of course, the corresponding many-body techniques are compatible with the standard techniques applicable to diagonal kernels~\cite{blaizot86}, i.e. setting $\Omega = 0$ in the expansion of off-diagonal kernels provides the same expansion for diagonal kernels as standard (symmetry-unrestricted) methods. The compact form given in Eq.~\ref{full2ndh} makes quite straightforward to see that one does indeed recover standard second-order (symmetry-unrestricted) MBPT for $\Omega=0$. Expanding the cluster amplitudes and using $U=U^{{\rm HF}}$ for simplicity leads to ${\cal T}^{\dagger (2)}_{ia}(0)=0$ for all $(a,i)$ and to
\begin{eqnarray}
h^{(2)}(0) &= & \sum_{i} t_{ii} + \frac{1}{2}\sum_{ij} \bar{v}^{\,\text{eff}}_{ijij}  - \frac{1}{4} \sum_{ijab}  \frac{|\bar{v}^{\,\text{eff}}_{ijab}|^2}{e_a+e_b-e_i-e_j} \,  \, , \label{full2ndhdiago}
\end{eqnarray}
along with ${\cal N}(0)=1$. This reduction to standard second-order MBPT originates from $\wp^{00}=0$ or equivalently to the fact that the bi-orthogonal system reduces to the mere eigenbasis of $H_0$ for $\Omega=0$, i.e. $\tilde{H}_{\text{eff}}(0)=H_{\text{eff}}$. 
Although the compact form of $h^{(2)}(\Omega)$ connects very naturally with $h^{(2)}(0)$, it hides the fact that its inherent complexity and richness would have forbidden to guess its form a priori, to obtain it from $h^{(2)}(0)$ via some sort of reverse engineering. As such, it was mandatory to develop a many-body expansion techniques of genuine {\it off-diagonal} kernels from which diagonal ones could be recovered as a particular case. Eventually, Eq.~\ref{full2ndhdiago} stresses that the diagonal kernel at play in SR calculations is explicitly correlated at the level of standard second-order MBPT. This expansion can be clearly extended to higher orders, at the price of complying with the associated increase of computational costs.

\section{Ab initio-driven EDF scheme}
\label{novelEDFscheme}

We thus propose to combine the formalism exemplified in the previous section with modern EDF techniques.
The resulting scheme can be summarized as follows, with the corresponding algorithm graphically illustrated in Fig.~\ref{algorithm}.

\subsection{Algorithm}
\label{algorithm}

\begin{enumerate}[A]
\item Reference state
\begin{enumerate}
\item Solve, e.g., symmetry-unrestricted Hartree-Fock equations in terms of $H_{\text{eff}}$ in the basis of interest to obtain the (deformed) reference state $| \Phi (0) \rangle$. This amounts to using 
\begin{equation*}
h^{(1)}(0) = h^{(1)}[\rho^{00}] \, ,
\end{equation*}
as an input diagonal functional. We denote by $N_b=N_h+N_p$ the dimension of the one-body Hilbert space, where $N_h$ denotes the number of occupied states of $| \Phi (0)\rangle$ and $N_p$ the number of unoccupied states.
\item Store single-particle energies $\{e_{\alpha}\}$ and wave-functions $\{\varphi_{\alpha}\}$.
\end{enumerate}
\item Single-reference calculations
\begin{enumerate}
\item Build the diagonal one-body density matrix $\rho^{00}$ along with the matrix elements of $T, V_{\text{eff}}, J_k$ and any other observable $O$ in the eigenbasis $\{\varphi_{\alpha}\}$ of $H_0$.
\item Compute from it the diagonal energy kernel at the chosen order $n$
\begin{equation*}
E_{{\rm SR}}\equiv  h^{(n)}(0) = h^{(n)}[\rho^{00}; \{e_{\alpha}\}] \, .
\end{equation*}
Proceed similarly for all the other observables $O$ of interest, i.e. compute
\begin{equation*}
O_{{\rm SR}}\equiv o^{(n)}(0) = o^{(n)}[\rho^{00}; \{e_{\alpha}\}] \, .
 \end{equation*}
\end{enumerate}
\item Multi-reference calculations
\begin{enumerate}
\item Discretize the intervals of integration over the three Euler angles $\Omega\equiv(\alpha,\beta,\gamma)$. 
\item For each combination of Euler angles
\begin{enumerate}
\item Build the $N_b \times N_b$ matrix $R_{\alpha\beta}(\Omega)\equiv \langle \alpha | R(\Omega) | \beta \rangle$ and its $N_h \times N_h$ reduction $M_{ij}(\Omega)$ to the subspace of hole states of $| \Phi (0) \rangle$. Compute the inverse $M^{-1}(\Omega)$.
\item Build the $N_p \times N_h$ rectangular matrix
\begin{equation*}
\wp^{0\Omega}_{ai}(\Omega)\equiv \sum_{i=1}^{N_h}R_{aj}(\Omega)M^{-1}_{ji}(\Omega) \, .
\end{equation*}
\item Build the bi-orthogonal bases according to Eqs.~\ref{rightbasis} and~\ref{leftbasis}. 
\item Transform the matrix elements of $T, V_{\text{eff}}, J_k$ and any other observable $O$ of interest into the bi-orthogonal system to generate the matrix elements of $\tilde{T}(\Omega)$, $\tilde{V}_{\text{eff}}(\Omega)$, $\tilde{J}_k(\Omega)$ and $\tilde{O}(\Omega)$, respectively.
\item Compute and store the off-diagonal linked/connected kernels at the chosen order $n$
\begin{eqnarray*}
h^{(n)}(\Omega) &=& h^{(n)}[\rho^{0\Omega}; \{e_{\alpha}\}] \, , \\
o^{(n)}(\Omega) &=& o^{(n)}[\rho^{0\Omega}; \{e_{\alpha}\}] \, , \\
j^{(n)}_{k=x,y,z}(\Omega) &=& j^{(n)}_{k=x,y,z}[\rho^{0\Omega}; \{e_{\alpha}\}] \, .
\end{eqnarray*}
\end{enumerate}
\item Using $j^{(n)}_{k=x,y,z}(\Omega)$ for the discretized values of the Euler angles, along with the initial condition ${\cal N}^{(n)}(0)=1$, integrate the three coupled ODEs (Eq.~\ref{NormkernelODE}) to obtain
\begin{eqnarray*}
{\cal N}^{(n)}(\Omega) &\equiv& \aleph^{(n)}[\mathbf{\rho}^{0\Omega} ; \{e_\alpha\}] \, \langle \Phi (0) | \Phi(\Omega) \rangle \, .
\end{eqnarray*}
for each combination of the Euler angles. 
\item Solve the Hill-Wheeler-Griffin equation to obtain the weights $f^{J}_K$ (Eq.~\ref{HWG}).
\item Calculate the energy of the yrast states through
\begin{equation*}
E^{J}_{{\rm MR}} = \frac{\sum_{K'K} f^{J\ast}_{K'} \, f^{J}_K  \int_{D_{SU(2)}} \! d\Omega \, D^{J \, \ast}_{K'K}(\Omega) \, h^{(n)}(\Omega) \, {\cal N}^{(n)}(\Omega)}{\sum_{K'K} f^{J\ast}_{K'} \, f^{J}_K  \int_{D_{SU(2)}} \! d\Omega \, D^{J \, \ast}_{K'K}(\Omega) \, {\cal N}^{(n)}(\Omega)}  \, .
\end{equation*}
Proceed similarly to compute other observables $O$ of interest\footnote{Care has to be taken for non-scalar operators such that the associated MR expression must be properly adapted from the one appropriate to the calculation of the energy~\cite{Yao:2014nta}.}.
\end{enumerate}
\end{enumerate}

\begin{figure}[t!]
\begin{center}
\includegraphics[clip=,width=0.5\textwidth,angle=0]{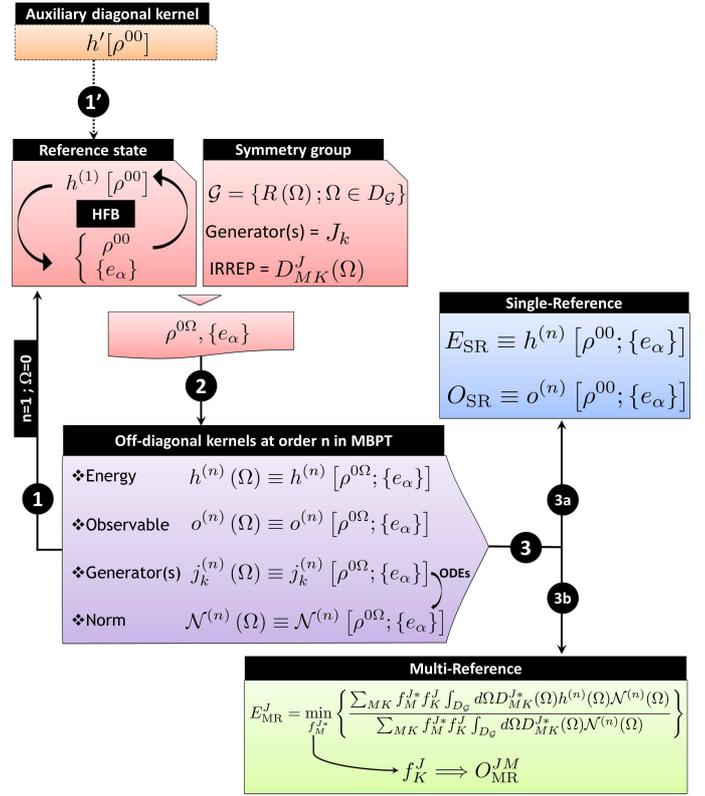} \\
\end{center}
\caption{
\label{algorithm}
(Color online) Algorithm at play for the ab initio-based EDF scheme.}
\end{figure}

\subsection{Discussion}
\label{Remarks}

A few further comments are in order regarding the EDF scheme we propose.
\begin{enumerate}
\item The proposed scheme is meant to formalize the EDF method as used in nuclear physics on a firmer ground than has been done in the past. It makes clear that the nuclear EDF method invokes fundamentally a {\it triple expansion} of the many-body solution. First, it involves an "horizontal" expansion in terms of non-orthogonal symmetry-breaking product states that provides an efficient way to resum so-called {\it non-dynamical}\footnote{{\it Dynamical} and {\it non-dynamical} refer to the wording at play in quantum chemistry.} correlations through the MR mixing. The associated off-diagonal kernels $N(\Omega)$ and $h(\Omega)$ are themselves "vertically" expanded in terms of particle-hole excitations, which corresponds to the treatment of so-called {\it dynamical} correlations. One merit of the many-body formalism developed in Ref.~\cite{duguet15a} is to propose, for the first time, a {\it consistent} simultaneous expansion of both dynamical and non-dynamical correlations, including their interference. Indeed, those two types of correlations are not orthogonal, i.e. not mutually exclusive, such that one may choose to convert one into the other. The dual expansion that lies at the core of the ab-initio-driven MR formalism is thus meant to optimize consistently the sharing between both kinds of correlations. Eventually, the scheme is implemented on the basis an operator $H_{\text{eff}}$ that must be seen as an effective {\it generator} of optimal EDF kernels at each given order of this dual expansion, i.e. its free parameters are to be optimized to reproduce experimental data at that chosen order. Whenever augmenting the explicit content of the kernels or/and of $H_{\text{eff}}$ itself, the parameters of the latter must be re-optimized fully. At this point, the form of $H_{\text{eff}}$ is to be guessed empirically. However, one can hope to design in the future a truly systematic {\it low-energy effective theory} that formulates the consistent expansion of the effective operator $H_{\text{eff}}$ in connection with the dual expansion of the many-body propagation to eventually achieve the triple expansion we alluded to above. In doing so, renormalizability of the kernels should guide the possible form of $H_{\text{eff}}$ at any given order in the many-body expansion, e.g. see Ref.~\cite{Moghrabi:2013fwa} for the discussion dedicated to the second-order diagonal kernel in isospin-symmetric infinite nuclear\footnote{Renormalizability implies in particular that the results are independent of the particular regularization scheme used.}. In such a low-energy effective theory, the generator $H_{\text{eff}}$ would formally converge to a realistic nuclear Hamiltonian $H$ at play in ab initio calculations whenever the many-body propagation is implemented to a sufficiently high order. 
\item The present scheme clarifies that building many-body corrections into the off-diagonal connected/linked energy kernel must be accompanied by a {\it consistent} enrichment of the off-diagonal norm kernel $N(\Omega)$. This is at variance with usual practice that consists of keeping the norm kernel to be the plain overlap between the two reference states involved independently of the many-body content of the energy kernel. This crucial point demonstrates that any well-founded formulation of the nuclear EDF method must consider energy and norm kernels on the same footing, which further stresses the necessity to focus on {\it off-diagonal} kernels. Indeed, the fact that the diagonal norm kernel can always be normalized to $1$, i.e. one can always redefine ${\cal N}(\Omega)$ in place of the original $N(\Omega)$, has undermined the necessity to understand how the norm kernel behaves as soon as the two reference states involved differ from one another. As for $SU(2)$ symmetry, the consistency between both kernels is ensured via the ODEs fulfilled by $N(\Omega)$ that involve off-diagonal kernels of the infinitesimal generators of the symmetry group computed at the same order as the energy kernel. For the one-dimensional $U(1)$ group, the norm kernel is related to the particle number operator kernel via one first-order ODE~\cite{duguet15b}. 
\item Mean-field expressions denote very restricted mathematical forms of the kernels, which happen to be functionals of the sole off-diagonal one-body density matrix, i.e.
\begin{eqnarray*}
h^{(1)}(\Omega) &\equiv& h^{(1)}[\mathbf{\rho}^{0\Omega}] \, , \\
o^{(1)}(\Omega) &\equiv& o^{(1)}[\mathbf{\rho}^{0\Omega}] \, , \\
j^{(1)}_k(\Omega) &\equiv& j^{(1)}_k[\mathbf{\rho}^{0\Omega}] \, , \\
{\cal N}^{(1)}(\Omega) &\equiv&  \langle \Phi (0) | \Phi(\Omega) \rangle \, .
\end{eqnarray*}
As soon as one includes explicit corrections to them, the kernels naturally become functionals of both the off-diagonal density matrix {\it and} the $N_b$ single-particle energies associated with the unperturbed Hamiltonian\footnote{Note that the single-particle energies are invariant under the rotation of $H_0$, i.e. under the transformation that leads from $|\Phi (0) \rangle$ to $|\Phi (\Omega) \rangle$. As such, they equally refer to $|\Phi (0) \rangle$ and $|\Phi (\Omega) \rangle$.}
\begin{eqnarray*}
h^{(n)}(\Omega) &\equiv& h^{(n)}[\mathbf{\rho}^{0\Omega} ; \{e_\alpha\}] \, , \\
o^{(n)}(\Omega) &\equiv& o^{(n)}[\mathbf{\rho}^{0\Omega} ; \{e_\alpha\}] \, , \\
j^{(n)}_k(\Omega) &\equiv& j^{(n)}_k[\mathbf{\rho}^{0\Omega} ; \{e_\alpha\}] \, , \\
{\cal N}^{(n)}(\Omega) &\equiv& \aleph^{(n)}[\mathbf{\rho}^{0\Omega} ; \{e_\alpha\}] \, \langle \Phi (0) | \Phi(\Omega) \rangle \, .
\end{eqnarray*}
Consequently, our proposal naturally extends to {\it off-diagonal} kernels the notion of orbital- and energy-dependent functionals at play in Density Functional Theory applicable to electronic systems~\cite{engel03a}.
\item Off-diagonal kernels are systematically improvable beyond traditional mean-field matrix elements, i.e. they are based on a systematic expansion. We advocate here to employ the simplest expansion based on low-order perturbation theory but other more involved choices could be envisioned, such as a low-order coupled-cluster truncation. Eventually, one aims at an optimal compromise between the complexity of the many-body formalism and that of the effective Hamiltonian, enabling the resulting method to be at the same time sufficiently rich and computationally manageable over the nuclear chart.
\item The reference state $|\Phi (0) \rangle$ is typically to be obtained on the basis of a mean-field-like kernel, e.g. the first-order diagonal kernel $h^{(1)}(0)$. Once this is done, the actual energy of the system can be implemented at a higher order $n$, both at the SR and MR levels. While a link is kept between both steps in the sense that both $h^{(1)}(0)$ and $h^{(n)}(\Omega)$ are generated from the same effective Hamiltonian $H_{\text{eff}}$, this demonstrates that the determination of the reference state is (can be) decoupled from the actual computation of the total energy. This is nothing but the logic typically followed in MBPT. As a matter of fact, one could go even further and choose to optimize the reference state on the basis of a kernel that is entirely disconnected from $h^{(n)}(\Omega)$, e.g. one could optimize $|\Phi (0) \rangle$ from an auxiliary, e.g. Skyrme, parametrization of the diagonal kernel  (see step (1') in Fig.~ \ref{algorithm}) that effectively account for some correlations before proceeding to the spuriosity-free and explicitly-correlated calculation of $E_{{\rm SR}}$ and $E^{J}_{{\rm MR}}$ on the basis of $h^{(n)}(\Omega)$. Such a scheme would require that the parameters of  $H_{\text{eff}}$ are adjusted {\it relative to} the auxiliary parametrization used to optimize $|\Phi (0) \rangle$ such that the parameter space effectively defining the approach would be larger than in the approach we presently advocate to follow.
\item Only at lowest order ($n=1$) can the symmetry-restored energy $E^{J}_{{\rm MR}}$ be factorized as the expectation value of $H_{\text{eff}}$ in a symmetry-projected wave-function~\cite{duguet15a}. As soon as higher orders are included ($n\geq 2$), this becomes impossible, i.e. the underlying symmetry-conserving wave-function is only {\it implicit}. This is actually necessary to obtain a connected/linked expression of the energy kernel, which is itself mandatory for the method to be size extensive, i.e. for the energy to scale correctly with particle number~\cite{shavitt09a}.
\item The present scheme offers a way to include many-body correlations consistently in the computation of all the observables of interest, i.e. not only for the energy but also for any observable $O$ such as, e.g., charge radii, electromagnetic moments etc.
\item Deriving the kernels from a true operator via explicit many-body techniques offers the possibility to carry out spuriousity-free, i.e. safe, MR-EDF calculations. It however does not guarantee it, i.e. an incautious truncation of the full many-body expansion may induce spurious self-interaction and self-pairing contributions~\cite{Bender:2008rn,duguet14a}. In the present discussion based on particle-number conserving reference states and properly antisymmetrized matrix elements, the procedure is always safe. When expanding more general off-diagonal kernels defined from Bogoliubov reference states~\cite{duguet15b}, more attention must however be paid to this question. Eventually, truncation schemes based on strict perturbation theory happen to be indeed safe.
\item To reduce the cost of the symmetry restoration~\cite{Bender:2005ri}, it will be of interest to apply the topological gaussian overlap approximation~\cite{onishi66,Hagino:2003dn} to the second-order off-diagonal kernels.
\end{enumerate}

\section{Conclusions and perspectives}
\label{concluandperspect}

This programmatic document lays down the possibility to build novel parametrizations of the {\it off-diagonal} energy and norm kernels that lie at the heart of the nuclear energy density functional method. The proposal is to exploit the ab initio many-body formalism recently proposed in Ref.~\cite{duguet15a} to guide the construction of safe, explicitly correlated and systematically improvable parametrizations. The many-body formalism of interest relies on the concepts of symmetry breaking {\it and} restoration that have made the fortune of the nuclear EDF method and is, as such, amenable to this guidance. We have detailed the proposal in its basic form, including the general equations and formulas that need to be implemented. In the mid-term future, we wish to investigate this scheme according to the following plan
\begin{enumerate}
\item Although $SU(2)$ and Slater determinants have been used in the present document as a prime example to expose the new EDF scheme, we plan to implement it first in connection with the breaking and the restoration of $U(1)$ (particle number) symmetry, which requires the handling of Bogoliubov reference states. The ab initio many-body method that can serve as a guidance and that will allow us to fully map the present discussion to the $U(1)$ group has been worked out in Ref.~\cite{duguet15b}.
\item Although this is a long-term objective, we will not exploit renormalizability as a guiding principle to build $H_{\text{eff}}$ at first. We will use two-body contact interactions containing both central and spin-orbit parts and regulated either with a Gaussian form factor or a sharp cut off.
\item The kernels will be computed at second order in perturbation theory. Although the computation of $n=2$ kernels is more costly than that of traditional mean-field ($n=1$) kernels, it is the least costly many-body method to correct for the deficiencies of the latter. As a matter of fact, there exist methods developed in quantum chemistry to reduce the corresponding cost from a scaling that naturally goes as $N_b^5$ down to a scaling that goes as, e.g., $N_b^3 \times N_h$~\cite{khoromskaia14a}. In considering these (approximation) techniques, it will be essential to only retain those that do not induce any pathology associated with spurious self-interaction and self-pairing processes~\cite{Bender:2008rn,duguet14a}, which are formally avoided in the first place in the proposed scheme. 
\end{enumerate}

The present discussion has focused on the restoration of symmetries, i.e. on multi-reference calculations that carry the mixing over the angle(s) $\Omega$ of the order parameter $g$ of the broken symmetry. This constitutes only one side of the coin of MR-EDF calculations that can/should also treat fluctuations over the norm $|g|$ of the order parameter. The possibility to formulate the mixing over $|g|$ on the same footing as what is proposed here for the symmetry restoration remains to be formulated. Although more challenging in several respect, the corresponding many-body formalism is currently being formulated~\cite{duguet15c} and will thus allow a consistent finalization of the presently proposed EDF scheme. Once this is done, it will be of interest to derive the corresponding (extended) random phase approximation. This can be achieved by taking the limit of the second-order off-diagonal kernels where $| \Phi^{(g')} \rangle$ and $| \Phi^{(g)} \rangle$ differ harmonically from a common reference state~\cite{jancovici64,brink68}.

\section*{Acknowledgments}


\end{document}